 \documentclass[doublespacing]{elsart}

\usepackage{amssymb}

\begin{document}

\begin{frontmatter}



\title{Stability of holonomic quantum computations}


\author[label1]{V.I. Kuvshinov}
\author[label2]{A.V. Kuzmin}
\address[label1]{Institute of Physics, 220072 Belarus, Minsk, Scorina,68.
                    Tel. 375-172-84-16-28, fax: 375-172-84-08-79.
                    E-mail: kuvshino@dragon.bas-net.by}
\address[label2]{Institute of Physics, 220072 Belarus, Minsk, Scorina,68.
                    E-mail: avkuzmin@dragon.bas-net.by}
\begin{abstract}
We study the stability of holonomic quantum computations with
respect to errors in assignment of control parameters. The general
expression for fidelity is obtaned. In the small errors limit the
simple formulae for the fidelity decrease rate is derived.
\end{abstract}

\end{frontmatter}

Holonomic approach to quantum computations was primarily proposed
in the Ref.~\cite{HQC}. Optical holonomic quantum computer was
proposed to realize this idea in a non-linear Kerr medium with the
degenerate states of laser beams to be interpreted as qubits and
logical gates to be realized by employing the existing devices of
quantum optics~\cite{OHQC}. Another implementation of holonomic
quantum computer based on trapped ions in optical cavity was
discussed recently from the viewpoint of its resilience to control
parameter errors (undesirable fluctuations of the laser beams
amplitudes)~\cite{ion}. Fidelity primarily proposed in the
Ref.~\cite{Peres} is widely used as a measure of the stability of
quantum systems. Fidelity close to unity means stability of
quantum system with respect to small perturbations of the
Hamiltonian (or more generally evolution operator)~\cite{Prozen1}.
Particulary it defines the resilience of quantum
computations~\cite{Prozen2}. Another approaches to determine the
stability of quantum systems were also proposed (for instance see
the Ref.~\cite{we}). It was demonstrated in particular for cavity
model of holonomic quantum computer that this realization of
logical gates provides fidelity and a success rate close to
unity~\cite{PRL}. In this Letter we derive the general expression
for the fidelity of any holonomic quantum computation. Taylor
expansion for fidelity as function of the control errors magnitude
for the general case of holonomic quantum computer is obtained in
the most relevant limit of small errors. For this purpose the
non-abelian Stokes theorem primarily utilized in connection with
the confinement problem of quarks and gluons~\cite{Simonov} is
used. The simple formulae for the fidelity decrease rate when
control errors grow is extracted. It is worth to note that
fidelity in particular for holonomic quantum computer is
mathematically similar to the Wilson loop defined in the theory of
strong interactions of elementary particles (quantum
chromodynamics), see the Ref.~\cite{confinement}. Namely, the
"area law" for Wilson loop means confinement of quarks and gluons,
as well the similar behavior of fidelity demonstrates the
instability of quantum system. This corresponds to the connection
between quantum chaos and confinement in the sense of the
Ref.~\cite{we}.

In holonomic quantum computer non-abelian geometric phases
(holonomies) are used for implementation of unitary
transformations (quantum gates) in the subspace $C^N$ spanned on
eigenvectors corresponding the degenerate eigenvalue of parametric
isospectral family of Hamiltonians \linebreak $F = \{H(\lambda) =
U(\lambda) H_0 U(\lambda)^+\}_{\lambda \in M}$~\cite{HQC}. The
$\lambda$'s are the control parameters and $M$ represents the
space of the control parameters. The subspace $C^N$ is called
quantum code ($N$ is the dimension of the degenerate computational
subspace). Quantum gates are realized when control parameters are
adiabatically driven along the loops in the control manifold $M$.
The unitary operator mapping the initial state vector into the
final one has the form $\bigoplus_{l=1}^R e^{i \phi_l}
\Gamma_{\gamma}(A_\mu^l)$, where $l$ enumerates the energy levels
of the system, $\phi_l$ is the dynamical phase, $R$ is the number
of different energy levels of the system under consideration and
the holonomy associated with the loop $\gamma \in M$ is given
by~\cite{PRA}:
\begin{equation}\label{Hol}
  \Gamma_{\gamma}(A_\mu) = \hat{P} \exp{i\int_{\gamma} A_{\mu} d \lambda_{\mu} }.
\end{equation}
Here $\hat{P}$ denotes the path ordering operator, $A_\mu $ is the
matrix valued adiabatic connection given by the expression:
\begin{equation}\label{A}
(A_\mu )_{mn} = \int d^3 x  \psi_m^* (x,\lambda) \left(
-i\frac{\partial}{\partial \lambda_\mu} \right) \psi_n
(x,\lambda),
\end{equation}
where integration goes over the spatial coordinates $x$ and
$\psi_n (x,\lambda), \quad n=\overline{1,N}$ are basis functions
of the corresponding eigenspace $C^N$. Dynamical phase will be
omitted bellow due to the suitable choice of the zero energy
level~\cite{HQC}. We shall consider the single subspace (no energy
level crossings are assumed).

Fidelity for holonomic quantum computer in the general case can be
written as follows:
\begin{equation}\label{fidelity}
  f = tr \left( \rho \Gamma_{\gamma^{\prime}}^{-1} \Gamma_{\gamma_0} \right) = tr \left( \rho \hat{P} \exp{\{
  i \int_{\delta \gamma} A_\mu d \lambda_\mu \} } \right)
  \end{equation}
where $\gamma_0$ is the adiabatic loop implementing the desirable
quantum gate, $\gamma^{\prime}$ is the actual loop with some
deviations from $\gamma_0$ due to practically unavoidable errors
in assignment of control parameters, $\rho$ denotes the density
matrix of the quantum state which stability is studied. The loop
$\delta \gamma$ is defined by the relation $\delta \gamma =
\gamma^{\prime -1} \cdot \gamma_0$. The $\cdot$-operation means
that to obtain the contour $\delta \gamma$ we primarily go over
the loop $\gamma_0$ in the straight direction and after that
travel along the loop $\gamma^{\prime}$ in the opposite direction.
For the definition and properties of this operation see
Ref.~\cite{HQC}. The contour $\delta \gamma$ is assumed to be
closed i.e. to form a loop.

Using the non-abelian Stokes theorem~\cite{Simonov} is it easy to
obtain the following expression for the fidelity (\ref{fidelity}):
\begin{equation}\label{FST}
  f = tr \rho \hat{P} \exp{\left( i \int_{\delta S} \left[\hat{P} e^{i
  \int_z^{x_0} A_\mu d \lambda_\mu} \right] F_{\chi \varrho} \left[ \hat{P} e^{i
  \int_{x_0}^z A_\nu d \lambda_\nu} \right] d \sigma_{\chi \varrho}(z) \right) }.
\end{equation}
Here $F_{\chi \varrho} = \partial_\chi A_\varrho -
\partial_\varrho A_\chi - i [A_\chi , A_\varrho ]_{-}$ is the
curvature tensor in the space of the control parameters,
$\partial_\chi \equiv \partial / \partial \lambda_\chi$, $[,]_{-}$
denotes the commutator, $\delta S$ is the surface spanned on the
loop $\delta \gamma$ and $d \sigma_{\chi \varrho}$ is the
projection of the infinitesimally small surface element of $\delta
S$ on the coordinate plane $\chi \varrho$ of the control parameter
space. In the formulae (\ref{FST}) and everywhere bellow the
summation over the indices of the curvature tensor $F_{\chi
\varrho}$ is performed under the condition $\chi > \varrho$.
Current point $z$ is located on the loop $\delta \gamma$ and point
$x_0$ is the arbitrary one. In the Ref.~\cite{Simonov} it was
demonstrated that when the non-abelian Stokes theorem is applied
the result does not depend on the particular form of $\delta S$
and the point $x_0$ is arbitrary if there are no monopole-like and
string-like topological structures. This is the case assumed here.
The general case deserves further investigation.

The general expression (\ref{FST}) is not convenient for the
practical use. To simplify it we consider more relevant limit of
small errors. Namely, we assume that the errors in the assignment
of control parameters $\delta \lambda_\mu$ are small and satisfy
the restrictions $|\delta \lambda_\mu | < ||A_\mu||^{-1}, \quad
|\delta \lambda_\chi \delta \lambda_\rho| < ||F_{\chi
\rho}||^{-1}$ ($\chi > \rho$). The connection and curvature tensor
are calculated in some point $\lambda_0$ defining the position of
the small loop $\delta \gamma$. The norm is defined as follows
$||B|| = sup \{ \sqrt{<\psi|B^+ B|\psi>}\left| \right. |\psi> \in
C^N, \quad <\psi|\psi> = 1 \}$. Here $B$ denotes both the
connection matrix components $A_\mu$ and the curvature tensor
matrix components $F_{\chi \rho}$ ($\chi > \rho$). Then the
following Taylor expansion in the point $\lambda_0 \in M$ can be
obtained for the fidelity (\ref{FST}):
$$
f =  1 + i tr \rho F_{\chi \varrho} \delta \lambda_\chi \delta
\lambda_\varrho -
  tr \rho [A_\mu , F_{\chi \varrho}]_{-} \delta \lambda_\mu \delta \lambda_\chi
 \delta \lambda_\varrho +  \left( i tr \rho A_\mu F_{\chi \varrho} A_\nu - \right.
$$
\begin{equation}\label{TaylorFidelity}
  \left. - \frac{i}{2}
 tr \rho [F_{\chi \varrho} , A_\mu A_\nu]_{+} - \frac{1}{2} tr \rho F_{\chi \varrho} F_{\mu \nu}
 \right) \delta \lambda_\chi \delta \lambda_\varrho \delta \lambda_\mu \delta
 \lambda_\nu + O(\delta \lambda^5).
\end{equation}
Here the point $\lambda_0$ defines the position of the small loop
$\delta \gamma$ in the space of parameters $M$ and $[,]_{+}$
denotes the anti-commutator. Under the stated restriction on the
control parameter errors higher order terms in
Eq.(\ref{TaylorFidelity}) give less significant contribution to
the fidelity then the terms having the lower order on $\delta
\lambda$. It is seen that the Taylor expansion for fidelity does
not have the linear on $\delta \lambda$ term. The cancellation of
the first order terms was noticed for the particular
implementation of holonomic quantum computer on trapped
ions~\cite{ion}. We have demonstrated that this conclusion does
not depend on particular realization of holonomic quantum
computation. The reason is that the difference between two vector
parallel transported along two different infinitesimally small
paths from one point to another is proportional to the area
enclosed by these contours and it does not depend on their
lengths.

Equality of the fidelity to the unity means that the computations
are stable. Namely, the result of computations is not changed,
when the quantum gate is perturbed. Expression
(\ref{TaylorFidelity}) shows that fidelity slightly deviates from
unity for small errors in assignment of control parameters. For
zero errors fidelity equals unity as it should be. The rate of
fidelity deviation from unity extracted from the second item of
the expansion (\ref{TaylorFidelity}) is determined by the
curvature tensor:
\begin{equation}\label{FidelityRate}
  \left| \frac{\delta f}{\delta S_{\mu \nu}} \right| = \left| tr \rho F_{\mu \nu}
  \right|
\end{equation}
where $\delta S_{\mu \nu} = \delta \lambda_\mu \delta \lambda_\nu$
is the "error area" in the plane $\mu \nu$. We see that the less
the curvature tensor leads the less the deviation of the fidelity
from unity and therefore calculations are more stable. This result
agrees with one of the Ref.~\cite{ion}, where for the particular
realization of holonomic quantum computer on trapped ions it was
demonstrated the one-qubit holonomic gate to be resilient to
control errors for large values of the squeezing parameter
(exponentially small values of the curvature tensor components).
From the physical point of view this result can be understood as
follows. The curvature tensor (as well as the area enclosed)
determines the difference between two vector parallel transported
along two different infinitesimally small paths from one point to
another. Therefore the less the curvature means the less the
difference between state vectors parallel transported along loops
$\gamma_0$ and $\gamma^{\prime}$. Thus the computation is more
stable. The condition for robust holonomic quantum computation is
$F_{\mu \nu} = 0$ for points belonging to the loop $\gamma_0$.

In conclusion, exploring the similarity between the mathematical
apparatus used in non-abelian gauge field theories and one used
for description of holonomic quantum computation we obtained the
general expression for fidelity of holonomic quantum computer. Its
Taylor expansion in the small control errors limit is derived and
the simple formulae for the rate of the fidelity deviation from
the unity is obtained. Our general results are in agreement with
ones obtained for the particular realization of holonomic quantum
computer on trapped ions. The formal mathematical analogy between
the fidelity and Wilson loop was noted. The question about the
relationship between the confinement of the strongly interacting
quarks and gluons and the stability of the corresponding systems
is under the investigation.

{\it Acknowledgements} This work has been supported by the grant
of the World Federation of Scientists, National Scholarship
Programme -- Belarus.

\end{document}